\documentclass[%
 reprint,
 amsmath,amssymb,
 aps,
 pra,
]{revtex4-2}

\usepackage{graphicx}
\usepackage{dcolumn}
\usepackage{bm}

\begin{document}

\preprint{APS/123-QED}

\title{Aharonov-Bohm effects on the GUP framework}

\author{Baoyu Tan}
\email{2022201126@buct.edu.cn}
\affiliation{College of Mathematics and Physics, Beijing University of Chemical Technology, 15 Beisanhuandonglu Street, Beijing, 100029, China.}

\date{\today}

\begin{abstract}
Modifying the fundamental commutation relation of quantum mechanics to reflect the influence of gravity is an important approach to reconcile the contradiction between quantum field theory and general relativity. In the past two decades, researchers have conducted extensive research on geometric phase problems in non-commutative spaces, but few have mentioned the correction of geometric phase problems using the Generalized Uncertainty Principle(GUP). This paper is the first to study the phase correction of Aharonov-Bohm(AB) effect by GUP.
\end{abstract}

\maketitle


\section{Introduction}

In recent decades, researchers have tried many different solutions to resolve the contradiction between general relativity and quantum theory. One attempt was to propose the theory of superstrings \cite{green1984superstring}, while the other attempted to reflect the influence of gravity by modifying the fundamental commutation relation of quantum mechanics. Non-commutative quantum mechanics \cite{connes1998noncommutative,seiberg1999string} and quantum mechanics under the Generalized Uncertainty Principle (GUP) \cite{kempf1995hilbert,kempf1997minimal,hossenfelder2006interpretation} are two major attempts in this regard.

In the past two decades, researchers have conducted extensive research on geometric phase problems in non-commutative quantum mechanics \cite{chaichian2001quantum,chaichian2002aharonov,falomir2002testing,mirza2004non,li2006aharonov,li2007topological,wang2007hmw}. Among them, the Aharonov-Bohm(AB) effect in non-commutative quantum mechanics is the most extensively studied \cite{aharonov1959significance}. In Ref. \cite{li2006aharonov}, the author applied Bopp shift to study the AB effect in non-commutative spaces. Ref. \cite{chaichian2001quantum,chaichian2002aharonov} propose a semi-classical effective Lagrangian to calculate the correction of AB effect by non-commutativity in space. Ref. \cite{ma2016time} studied the AB effect in non-commutative spaces using Seiberg-Witten mapping. But up to now, no one has studied the AB effect of GUP correction. This paper uses the methods in Ref. \cite{ali2009discreteness,das2010discreteness} to calculate the phase correction in the AB effect within the framework of GUP.

\section{GUP and deformed Dirac equation}

According to Ref. \cite{ali2009discreteness,das2010discreteness}, we have a GUP that is consistent with DSR theory, string theory, and black hole physics, and satisfies $[x_i,x_j]=[p_i,p_j]=0$(via the Jacobi identity):
\begin{equation}
[x_i,p_j]=i\hbar\left[\delta_{ij}-a(p\delta_{ij}+\frac{p_ip_j}{p})+a^2(p^2\delta_{ij}+3p_ip_j)\right].\label{eq:1}
\end{equation}
\begin{align}
\Delta x\Delta p&\geq\frac{\hbar}{2}[1-2a\langle p\rangle+4a^2\langle p^2\rangle]\nonumber\\
&\geq\frac{\hbar}{2}\left[1+\left(\frac{a}{\sqrt{\langle p^2\rangle}}+4a^2\right)\Delta p^2+4a^2\langle p\rangle^2-2a\sqrt{\langle p^2\rangle}\right].
\end{align}
Where $a=a_0/M_{pl}c=a_0l_{pl}/\hbar$ is GUP parameter, $M_{pl}$ is Planck mass, $l_{pl}\approx 10^{-35}m$ is Planck length, $M_{pl}c^2\approx 10^19GeV$ is Planck energy, usually assuming $a_0\approx 1$. Due to the DSR transformation not only retaining the speed of light, but also the Planck momentum and Planck length. So there are the following minimum measurable length and maximum measurable momentum:
\begin{equation}
\Delta x\geq(\Delta x)_{\min}\approx a_0l_{pl}.
\end{equation}
\begin{equation}
\Delta p\leq(\Delta p)_{\max}\approx \frac{M_{pl}c}{a_0}.
\end{equation}
To satisfy Eq. (\ref{eq:1}), we define the following coordinates and momentum:
\begin{equation}
x_i=x_{0i},~~~~p_i=p_{0i}(1-ap_0+2a^2p_0^2).
\end{equation}
Where $x_{0i}$, $p_{0j}$ are uncorrected coordinates and momentum, satisfying the commutation relation $[x_{0i},p_{0j}]=i\hbar\delta_{ij}$, $p_{0i}=-i\hbar\partial/\partial x_{0i}$, $p_0=\sqrt{p_{0x}^2+p_{0y}^2+p_{0z}^2}$.

We linearize $p_0$ and make substitution $p_0\to \vec{\alpha}\cdot\vec{p}$. Where $\alpha_i(i=1,2,3)$ and $\beta$ are the Dirac matrices:
\begin{eqnarray}
\alpha_i=\left(
\begin{array}{cc}
0 & \sigma_i \\
\sigma_i & 0
\end{array}\right),~~~~\beta=\left(
\begin{array}{cc}
I & 0 \\
0 & -I
\end{array}\right).
\end{eqnarray}
The Dirac equation corrected by GUP can be written to O(a) order as:
\begin{align}
H\psi(\vec{r})=&(c\vec{\alpha}\cdot\vec{p}+\beta mc^2)\psi(\vec{r})\nonumber\\
=&[c\vec{\alpha}\cdot\vec{p_0}+ca(\vec{\alpha}\cdot\vec{p_0})(\vec{\alpha}\cdot\vec{p_0})+\beta mc^2]\psi(\vec{r}).\label{eq:2}
\end{align}
Write Eq. (\ref{eq:2}) in the form of Lorentz covariance and use the natural unit system throughout the following text. Define $\gamma^0=\beta$, $\gamma^i=\beta\alpha^i(i=1,2,3)$:
\begin{equation}
(i\gamma^\mu p_\mu-m)\psi=0.\label{eq:3}
\end{equation}
Consider GUP correction of Eq. (\ref{eq:3}), accurate to order O (a), we define:
\begin{equation}
x^\mu=x_0^\mu,~~~~p_\mu=p_{0\mu}(1-a\gamma^\mu p_{0\mu}).
\end{equation}
Where $x_0^\mu$ and $p_{0\mu}$ are the coordinates and momentum without considering GUP correction.

\section{GUP corrected AB effects}

Now let's consider the phase correction of GUP on AB effect. The Lagrangian describing a relativistic spin-half charged particle with under the electromagnetic field background modified by GUP is:
\begin{equation}
\mathcal{L}^{\text{GUP}}=\bar{\psi}[\gamma^\mu p_{0\mu}(1-a\gamma^\mu p_{0\mu})-q\gamma^\mu A_\mu-m]\psi.
\end{equation}
In order to make the Lagrangian invariant under gauge transformations, we introduce the covariant derivative $D_\mu$. The Lagrangian can be rewritten as:
\begin{equation}
\mathcal{L}^{\text{GUP}}=\bar{\psi}(i\gamma^\mu D_\mu^{\text{GUP}}-m)\psi.
\end{equation}
The covariant derivative without considering GUP correction is $D_\mu=\partial_\mu+iqA_\mu$, and covariant derivative after considering GUP correction is:
\begin{equation}
D_\mu^{\text{GUP}}=\partial_{0\mu}(1-a\gamma^\mu\partial_{0\mu})+iqA_\mu.
\end{equation}
We obtained the dynamical equation of the AB effect with GUP correction accurate to O(a) order:
\begin{equation}
(i\gamma^\mu D_\mu^{\text{GUP}}-m)\psi=0.\label{eq:4}
\end{equation}
Comparing Eq. (\ref{eq:4}) with the dynamical equation without GUP correction, we can find that there is an additional term with GUP parameter.

Starting from dynamic equation (\ref{eq:4}), we can obtain the GUP corrected AB phase as follows:
\begin{equation}
\phi^{\text{GUP}}=\phi+\delta\phi.
\end{equation}
The term that has not been corrected by GUP is a well-known result of general quantum mechanics:
\begin{equation}
\phi=q\oint A_\mu\mathrm{d}x^\mu.
\end{equation}
And the correction term for GUP is:
\begin{equation}
\delta\phi=-aq\oint\gamma^\nu p_{0\nu}p_{0\mu}\mathrm{d}x^\mu.
\end{equation}
Obviously, when $a\to0$, the correction term $\delta\phi$ of GUP disappears, returning to the results of general quantum mechanics that we are familiar with.

\section{Conclusions}

We use the GUP modified Dirac equation to provide GUP correction for phase in AB effect that depends on GUP parameters. When the GUP parameter approaches zero, it can return to the general quantum mechanics situation. Our method can be easily extended to other geometric phase problems corrected by GUP, such as AC effect, HMW effect, and Anandan phase.

\section*{Acknowledgments}

The authors would like to thank Prof. Jian Jing and his student LiuBiao Ma and Zheng Wang from the Department of Physics, Beijing University of Chemical Technology for their valuable comments and suggestions during the completion of this manuscript.

\nocite{*}

\bibliography{GUP-AB-effect}

\end{document}